\documentclass[aps, prd, twocolumn, notitlepage, nofootinbib]{revtex4-1}
\usepackage{graphicx}
\usepackage{float}
\usepackage{amsmath}
\usepackage{color}
\usepackage{tensor}
\usepackage{epstopdf}
\usepackage{xcolor}
\usepackage{wasysym}
\usepackage{hyperref}
\hypersetup{
    colorlinks,
    linkcolor={blue!50!black},
    citecolor={red!50!black},
    urlcolor={blue!80!black}
}

\usepackage{multibib}

\newcommand{\be}{\begin{equation}}
\newcommand{\ee}{\end{equation}}
\newcommand{\ba}{\begin{eqnarray}}
\newcommand{\ea}{\end{eqnarray}}

\newcommand{\beq}{\begin{equation}}
\newcommand{\eeq}{\end{equation}}
\newcommand{\beqa}{\begin{eqnarray}}
\newcommand{\eeqa}{\end{eqnarray}}
\newcommand{\nn}{\nonumber}

\begin{document}

\title{Einstein Quartic Gravity: Shadows, Signals, and Stability}

\author{Hossein Khodabakhshi}
\email{h.khodabakhshi@ipm.ir}
\affiliation{Department of Physics, University of Tehran, 14395-547 Tehran, Iran,}
\affiliation{School of Particles and Accelerators, Institute for Research in Fundamental Sciences (IPM), 19395-5531 Tehran, Iran}
 \affiliation{Department of Physics and Astronomy, University of Waterloo, Waterloo, Ontario, Canada N2L 3G1}
 \author{Andrea Giaimo}
\email{giaimo.andrea.93@gmail.com}
\affiliation{Physics Department, University of Calabria, I-87036 Arcavacata di Rende, Cosenza, Italy}
\affiliation{Department of Physics and Astronomy, University of Waterloo, Waterloo, Ontario, Canada, N2L 3G1}
\author{Robert B. Mann}
\email{rbmann@uwaterloo.ca}
\affiliation{Department of Physics and Astronomy, University of Waterloo, Waterloo, Ontario, Canada, N2L 3G1}
\affiliation{Perimeter Institute, 31 Caroline St. N., Waterloo, Ontario, N2L 2Y5, Canada}

%


\begin{abstract}
 Using a continued fraction ansatz we obtain an analytic approximation for a spherically symmetric black hole solution to Einsteinian Quartic Gravity (EQG), the next simplest Generalized Quasi-Topological Gravity (GQTG) after Einsteinian Cubic Gravity (ECG). This approximate solution is valid everywhere outside of the horizon and we use it to investigate the orbit of massive test bodies near a black hole, specifically computing the innermost stable circular orbit. Using Shapiro time delay we calculate the constraints on the EQG coupling parameter. Finally we compute the shadow of an EQG black hole and figure out it to be larger than its Einsteinian counterpart in general relativity for the same value of the mass. By applying our results to Sagittarius A* (Sgr A*) at the center of Milky Way we find, similar to ECG black holes, that departures from general relativity are small but distinguishable for EQG black holes.
\end{abstract}

\maketitle

\onecolumngrid


\section{Introduction}

General relativity continues to enjoy a spectacular degree of empirical success.  In the previous century a broad array of Earth-based and solar system tests repeatedly indicated that it was the preferred theory relative to other competitors \cite{Will:2014kxa}.  In the present century both LIGO observations 
\cite{Abbott2019}  and the Event Horizon Telescope \cite{EHT} have provided us with the first direct tests  in strong gravitational fields, with all observations fully in accord with  general relativity.

Nevertheless investigations of competitors to general relativity have continued apace.  This is driven by a number of considerations, the most prominent being that general relativity remains stubbornly resistant to quantization.
Although a quantum theory of gravity still eludes us, it is possible to derive some generic theoretical expectations of what such a theory might produce, and it is conceivable that these could have implications in astrophysics and cosmology.

One generic consequence of quantizing gravity appears to be the inclusion of  higher curvature corrections. 
Inclusion of terms in the action that are quadratic in the curvature can lead to a renormalizable theory of quantum gravity~\cite{Stelle}. Furthermore low energy effective actions derived from string theory generally yield various higher-derivative gravitational theories~\cite{Zwiebach,Metsaev,Myers87}.  Within  the AdS/CFT correspondence, 
higher curvature theories have been useful as (at least) toy models that allow us 
to make contact with a wider class of CFTs and to investigate effects beyond the large $N$ limit~\cite{Kats,Myers10,Hofman,Myers11,Myers08}.

However   most higher curvature theories have negative energy excitations (or ghosts)~\cite{Deser} when
linearized about constant curvature backgrounds, and so attention has concentrated on those that have such ghost degrees of freedom in their propagator~\cite{Modesto,Biswas1,Biswas2}.   The Lovelock class of theories~\cite{Lovelock:1971yv} are known to be the most general class  that are ghost-free on any background.  However
curvature terms of order $k$ in the action in this class of theories are topological invariants in $d = 2k$ dimensions, vanishing identically for $d < 2k$.  A recent generalization of this class, known as   Quasi-topological gravity 
\cite{Myers:2010ru, Oliva:2010eb, Cisterna:2017umf} provide additional examples of higher curvature theories that
have  yielded some interesting results in the context of holography~\cite{Myers:2010jv, Myers:2010xs}.  
However they are also trivial in four dimensions.

Within the last few years a new class of higher-curvature theories have been discovered that 
are neither topological nor trivial in four dimensions and that have the same graviton spectrum as general relativity on constant curvature backgrounds. Known as \textit{Generalized Quasi-Topological Gravity} (GQTG)~\cite{Pablo1,Hennigar:2017ego, Ahmed:2017jod}, their equations of motion for static, spherically symmetric spacetimes are sufficiently simple  to allow for a non-perturbative study of black hole solutions.  The first such theory to be
explored was cubic in curvature, and is known as {\it Einsteinian Cubic Gravity} (ECG)~\cite{Pablo1}.  A set of quartic theories was obtained shortly afterward~\cite{Ahmed:2017jod}, and recently it was 
shown~\cite{Bueno:2019ycr}  that GQTG can be constructed for arbitrarily high powers of the curvature  and in general dimensions. Recursive formulas were derived that allow one to  systematically construct  $n$-th order curvature densities  from lower order ones, as well as explicit expressions valid at any order.

A salient feature of GQTG theories is that their  field equations admit static, spherically symmetric (SSS) solutions with a single metric function,
\beq\label{metsss}
ds^2=-f(r) dt^2+\frac{dr^2}{f(r)}+r^2d\Omega^2_{(2)}  
\eeq
naturally generalizing the Schwarzschild solution. Apart from the technical convenience of simplifying  the study of black holes, this non-trivial feature is also responsible for the absence of ghosts and integrability.  There is a single field for $f(r)$ that is a total derivative, yielding a   non-linear second order differential equation upon integration, with the integration constant being related to the mass~\cite{Bueno:2017sui,Bueno:2019ycr}.  It is this  integrability that allows exact, analytic investigations of  black hole thermodynamics, despite the lack of exact solutions\footnote{For a special critical limit, certain exact solutions can be found~\cite{Feng:2017tev}.}  to the field equations~\cite{Robie17,Bueno:2016lrh}. These studies have revealed that small, asymptotically flat black hole solutions become stable, a result that may have implications in light of the information loss problem~\cite{Bueno:2017qce}. Studies of the thermodynamics of AdS black branes have revealed novel phase structure, suggesting this class of theories will provide rich holographic toy models~\cite{Hennigar:2017umz}, and a thorough study 
of black hole thermodynamics has been carried out for the cubic \cite{Mir:2019ecg} and quartic \cite{Mir:2019rik} versions of GQTG.

GQTGs are non-trivial in four dimensions.  No dimensional  reduction is required   to interpret solutions to
a GQTG theory.  In conjunction with the properties noted above, we regard  this  as sufficient reason to embark on a program to investigate the compatibility of GQTGs with observational tests.  The first steps along these lines were taken with ECG~\cite{Hennigar:2018hza}, in which  constraints that arise from solar system tests and potential signatures from black hole shadows were computed that could be constrained by the Event Horizon Telescope (EHT)~\cite{EHT}.  For the  largest value of the ECG coupling constant permitted by  Shapiro time delay, the angular radius of a shadow from a non-rotating black hole is enlarged by about  5 parts per million as compared to general relativity -- a small effect not distinguishable with current technology.  However a further
study~\cite{Poshteh:2018wqy} indicated that angular positions of gravitationally lensed images in ECG could
deviate from general relativity by as much as milliarcseconds, suggesting observational tests of ECG are indeed feasible.

In this paper we consider basic phenomenological tests of Einstein Quartic Gravity (EQG), the next simplest GQTG after ECG.  The EQG class of theories all have actions quartic in the curvature and were obtained shortly after the construction of ECG \cite{Ahmed:2017jod}.  There are six such quartic curvature combinations  that are nontrivial in (3+1) dimensions, leading to the introduction of six new coupling constants. However the imposition of spherical symmetry yields a degeneracy insofar as their field equations  differ by terms that vanish for a static spherically symmetric (SSS) metric.  We shall consider phenomenological implications of EQG under this ansatz, thereby obtaining constraints on a linear combination of the six couplings.

We pause to comment that the EQG theory we are considering is a particular example of a quartic GQTG and it does not meet the 
criteria used in finding ``Einsteinian" higher-curvature theories \cite{Bueno:2016ypa}.  Although the conditions satisfied by 
 GQTG theories and the more general conditions for the Einsteinian theories coincide for ECG\footnote{For this reason `Einsteinian Cubic Gravity' and 
 `Einstein Cubic Gravity' refer to the same theory.}, this is not true in general.  
However  there is a particular quartic GQTG that also meets the Einsteinian criteria  \cite{Bueno:2016ypa}, and we expect that some combination of 
the EQG  invariants we consider (perhaps with some possibly trivial densities) could satisfy these other criteria.  As we will see that in the spherically symmetric case  there is only a single parameter due to the aforementioned degeneracy, we shall not pursue this issue further.

We likewise emphasize that the EQG theories we consider  are quartic generalized quasi-topological gravities, and  
 should be distinguished from a broader set of quasitopological quartic theories 
 constructed so that they share the spectrum of Einstein gravity when linearized on a maximally symmetric background \cite{Bueno:2016ypa}.  EQGs are constructed by requiring that there is a single independent field equation for only one metric function under the restriction of spherical symmetry; when this is satisfied EQGs also have the same graviton spectrum and match the linearized Einstein equations on a constant curvature background up to a redefinition of Newton's constant \cite{Ahmed:2017jod}.   Although  recursive formulas have been
 obtained that  allow for the systematic construction of actions that are $n$-th order in curvature  from lower order ones \cite{Bueno:2019ycr}    EQGs are of particular interest because  
 they have the highest degree of curvature  possible that allows for an analytic solution 
 of the near horizon equations for the temperature and mass in terms of the horizon radius $r_+$.  
 
One technical challenge in studying EQG is that an analytic solution is not readily available. While numerical solutions can be obtained, we find it more productive to employ a continued fraction ansatz.  This was shown to yield a highly accurate analytic approximate solution to the field equations for ECG~\cite{Hennigar:2018hza}, and we find the same to be true for EQG as well. This approach has been successfully applied  in a variety of contexts~\cite{Rezzolla:2014mua, Kokkotas:2017zwt, Kokkotas:2017ymc, Konoplya:2016jvv}, and we expect it will have use  in future investigations (for example with  quasi-normal modes) as well.  We shall use  the continued fraction solution to analyze solar system tests, the motion of particles around a black hole in EQG, and the properties of a black hole shadow~\cite{synge, Bardeen, Vries, Bambi1,Eiroa1,Eiroa2,Eiroa3,Bambi2, Grenzebach}). 
 
We show that EQG is compatible with solar system tests for relatively large values of the coupling.  In particular, from 
 Shapiro time delay we find the strongest  constraint on the EQG coupling constant $K$ provided by solar system tests. Furthermore, we find that the radius of the innermost stable circular orbit (ISCO) around an SSS EQG black hole and the angular momentum of a test body at this radius increases with increasing $K$ as compared to their corresponding values in general relativity.  

Moving on to investigate null geodesics around an EQG black hole, we find that  its photon shadow is enlarged compared to  its non-rotating counterpart in general relativity.  We apply our results to the supermassive black hole Sagittarius A* (Sgr A*) at the center of our Galaxy and show,  similar to the ECG case, that the angular radius of the shadow  increases with increasing $K$ by an amount enticingly close to what could be experimentally detected, consistent with solar system tests.  This suggests that  important constraints on $K$ and on EQG in general
could be provided (at least in principle)  from EHT observations.

Our paper is organized as follows. In section~\ref{two} we review the near horizon,  asymptotic, and numeric solutions. We then compute the continued fraction expansion in the next section, obtaining an approximate analytic solution
in the SSS case.  In section~\ref{sec:BH_props} we investigate various properties of an  EQG black hole and   orbits of massive particles around it. In section~\ref{sec:testEQG} using Shapiro time delay we constrain the coupling constant of EQG. We study the null geodesics in EQG  in section~\ref{sec:shadow} and present our results for Sgr A* shadow and in the last section  A number of useful results are summarized in the appendices. We conclude our paper with a discussion of the phenomenological prospects of EQG.  We work in units where $G = c= 1$.

\section{Spherically Symmetric Solution in EQG}
\label{two}
The action for EQG is
\beq \label{actionEQG}
S = \frac{1}{16\pi} \int d^4 x \sqrt{-g} \left[R - \sum_{i=1}^{6} \hat{\lambda}_{(i)} \mathcal{S}_{4}^{(i)}\right],
\eeq
where $R$ is the usual Ricci scalar and $\mathcal{S}_{4}^{(i)}$ are called \textit{quasi-topological Lagrangian densities}, whose analytical expressions are given in the Appendix \ref{A}. 
We restrict ourselves to asymptotically flat, static, and spherically symmetric vacuum black holes, whose metric is 
given by \eqref{metsss}, with  $\lim_{r\to\infty}f(r) = 1$.  
In this case, the only independent field equation is
\begin{align} \label{eqn:feq}
&-(f-1)r - \frac{24}{5} K \bigg[\frac{1}{r^2} ff'f''(f-1-\frac{1}{2}rf') + \frac{1}{8r} f'^4 + \frac{1}{6 r^2}f'^3(f+2)+\frac{1}{r^3} f f'^2 (1-f)\bigg] = 2 M,
\end{align} 
with a prime denoting differentiation with respect to $r$.   We see the remarkable property of EQG (shared by all GQTG theories) that the SSS field equations reduce to a single 2nd-order differential equation for one metric function.

The imposition of spherical symmetry yields a degeneracy amongst the different theories in \eqref{actionEQG} 
in that the constant $K$ is a linear combination of
the six EQG coupling constants
\be\label{EQGK}
K\equiv -\frac{5}{6} \left(\sum_{i=1}^{6} \lambda_{(i)} \right) \, ,
\ee
where we find it convenient to write 
\begin{align}
{\lambda}_{(1)} &= -\frac{6}{5} \hat{\lambda}_{(1)}\,, \; \; \; {\lambda}_{(2)} = -3\hat{\lambda}_{(2)}\,, \; \; \; {\lambda}_{(3)} = -\frac{12}{5} \hat{\lambda}_{(3)}\,, \; \; \; {\lambda}_{(4)} = -\frac{24}{5} \hat{\lambda}_{(4)}\,, \; \; \; {\lambda}_{(5)} = -\frac{24}{5} \hat{\lambda}_{(5)}\,, \; \; \; {\lambda}_{(6)} = -\frac{96}{5} \hat{\lambda}_{(6)}\,.
\end{align}
The combination \eqref{EQGK} appears because each term $\mathcal{S}_{4}^{(i)}$ gives the same contribution to the field equation~\cite{Ahmed:2017jod}. This degeneracy means that there is one parameter from EQG that be constrained empirically from analysis of this class of solutions.

The quantity $M$ appearing on the right-hand side of the equation is the ADM mass of the black 
hole~\cite{Bueno:2016lrh, Hennigar:2017ego}. We shall see below that  asymptotic flatness requires $K > 0$ in what follows.   We will solve the field equation in two different regions of the spacetime: the near-horizon region and the large-$r$ region and then we will present a continued fraction expansion that provides an accurate and convenient approximation of the solution everywhere outside of the horizon.

\subsection{Near-horizon region}

As we shall be interested in black hole solutions, we begin by solving the field equations near the horizon via a series expansion using the ansatz  
\be\label{eqn:nh_expand} 
f_{\rm nh}(r) = 4 \pi T (r-r_+) + \sum_{n=2}^{n = \infty} a_n (r-r_+)^n \, ,
\ee 
ensuring that the metric function $f(r)$ vanishes linearly at the horizon ($r=r_+$), with $T = f'(r_+)/4\pi$  the Hawking temperature. We then obtain
\be\label{eqn:mass_equat}
M= \frac{1}{2} r_+ -\frac{128 K}{5} \frac{(\pi T)^3}{r_+^2} (3 \pi r_+ T +2) \,,
\ee
and
\be\label{eqn:temp_equat}
\frac{256 K}{5} \frac{\pi^4}{r_+^2} \; T^4 + \frac{512 K}{5} \frac{\pi^3}{r_+^3} \; T^3 + (4\pi T r_+ -1) = 0 \,
\ee
by substituting \eqref{eqn:nh_expand} into the field equations~\eqref{eqn:feq}. 

We therefore can 
explicitly obtain the mass and temperature of the black hole in terms of its horizon radius $r_+$ and the coupling functions,  even though we do not have an explicit expression for $f(r)$.
Defining the following quantities in terms of $r_+$ and the coupling $K$
\be
\tau \equiv 16 - \frac{20}{(25K)^\frac{1}{3}} r_+^2 + \frac{(25K)^\frac{1}{3}}{K} r_+^4 \,,
\ee
and 
\be
\xi \equiv \sqrt{ 48 + \frac{128}{\sqrt{\tau}} + \frac{10}{K \sqrt{\tau}} \, r_+^6 - \tau} \,,
\ee
we can solve~\eqref{eqn:mass_equat},~\eqref{eqn:temp_equat} 
\begin{align}
T &=\frac{1}{4 \pi r_+} \left[ \frac{1}{2} (\xi - \sqrt{\tau})-2 \right]\,,
\nn\\
M &=\frac{1}{2} r_+ \Bigl(-\frac{2048 K }{5 r_+^6} - 20 \Bigr) + \sqrt{\tau} \Bigl( -\frac{32 K}{r_+^5} + \frac{2 (25K)^\frac{1}{3}}{5r_+} -\frac{3r_+}{4} - \frac{8 K}{(25K)^\frac{1}{3} r_+^3} \Bigr) 
\nn\\
&+\frac{1}{\sqrt{\tau}} \Bigl( -\frac{1536 K}{5 r_+^5} - 24 r_+ \Bigr) + \xi \Bigl(  \frac{128K}{5 r_+^5} -\frac{8K}{(25K)^\frac{1}{3} r_+^3} + \frac{2(25K)^\frac{1}{3} }{5r_+} +\frac{3r_+}{4} \Bigr) 
\nn\\ 
&+\frac{\xi}{\sqrt{\tau}} \Bigl( \frac{128K}{5 r_+^5} + 2r_+ \Bigr) + (\xi \sqrt{\tau}) \frac{24K}{5 r_+^5}  
\label{eqn:mass_temp}
\end{align}
for the temperature and mass.   

We also find that the field equations~\eqref{eqn:feq} do not determine the parameter $a_2$ in the expansion
\eqref{eqn:nh_expand}. However all remaining $a_n$ for $n > 2$ are determined by (rather large) expressions involving $K$, $T$, $r_+$, and $a_2$.  

\subsection{Large-r asymptotic region}\label{seclgr}

We  consider next the large-$r$ asymptotic region. To obtain an approximate solution we linearize the field equations about the Schwarzschild background:
\be 
f_{\rm asymp} = 1 - \frac{2 M}{r} + \epsilon h(r),
\ee
where the field equations determine $h(r)$. Retaining terms  only to order $\epsilon$, the resulting differential equation for $h(r)$ takes the form
\be 
h'' + \gamma(r) h' + \delta (r) h = g(r),
\ee
where
\begin{align}
\gamma(r) &= - \frac{8M-5r}{(2M - r) r} \, , 
\nn\\
\delta(r) &=  \frac{86 \left[ - M^2  K \left(M-\frac{9r}{43}\right) - \frac{5}{1376}r^9 \right]}{-9 r^2 K M^2 (2M-r)}  \, ,
\nn\\
g(r) &=  \frac{M(-54r +97M)}{9r^3(2M-r)} \, .
\end{align}
In the large $r$ limit, the homogenous equation reads
\be 
h_h'' - \frac{5}{r} h_h' - \omega ^2 r^6 h_h = 0 \, ,
\ee
where the parameter $\omega$ is defined by
\be
\omega^2 \equiv \frac{5}{144 K M^2} \, ,
\ee
and it can be solved exactly in terms of Bessel functions:
\begin{align} 
h_h &= r^3 \bigg[\tilde{A} I_{\frac{3}{4}}\left(\frac{\omega r^4}{4}\right) + \tilde{B}  K_{\frac{3}{4}}\left(\frac{ \omega r^4}{4} \right)\bigg],
\end{align}
where $I_\nu(x)$ and $K_\nu (x)$ are the modified Bessel functions of the first and second kinds, respectively. To leading order in large $r$, this can be expanded as
\be\label{eqn:asymp_homog} 
h_h(r) \approx A r \exp\left[ \frac{\omega r^4}{4}\right] + B r  \exp\left[ - \frac{\omega r^4}{4}\right], 
\ee
absorbing various constants into the definitions of $A$ and $B$ (compared to $\tilde{A}$ and $\tilde{B}$).  We see that the homogenous solution  consists of a growing mode and a decaying mode. Asymptotic flatness demands that we set $A = 0$, while the second term decays super-exponentially and can therefore be neglected.\footnote{This assumes that $\omega^2 > 0$, in turn requiring $K > 0$. In cases where $K < 0$, the homogeneous solution contains oscillating terms that spoil the asymptotic flatness.  The only viable solution in this case is to set the homogenous solution to zero.}

There is also a particular solution
\begin{align}
h_{\rm p} &= - \frac{864}{5} \frac{K M^3}{r^9} + \frac{1552}{5} \frac{K }{r^{10}} + O \left( \frac{K^2 \, M^5} {r^{17}} \right),
\end{align}
that is more relevant as it implies
\be 
f(r) \approx 1 - \frac{2 M}{r} + h_{\rm p}  
\ee
since it clearly dominates over the super-exponentially decaying  homogenous solution at large $r$.

\section{Continued fraction approximation}

Neither the near horizon approximation nor the asymptotic solution are accurate in the entire space-time  outside the horizon. To complete the solution we can solve equations of motion in the intermediate regime numerically.  To do so, we choose a value for the free parameter $a_2$ for a given choice of $M$ and $K$. Using values of the near horizon expansion we write 
\begin{align}
f(r_{+}+\epsilon)=4 \pi T \epsilon + a_2 \epsilon^2 \nonumber\\
f'(r_{+}+\epsilon)=4 \pi T + 2a_2 \epsilon,
\label{df}
\end{align}
for the initial data for the differential equation just outside the horizon,
where $\epsilon$ is a small, positive quantity.  Noting (\ref{eqn:asymp_homog}),  a generic choice of $a_2$ excites the exponentially growing mode. To get the asymptotically flat solution $a_2$ must be chosen carefully with high precision. For some large values of $r$ (compared with the other scales in the problem) we can obtain a satisfactory numerical solution  consistent with the asymptotic expansion to a high degree of accuracy. There is a unique value of $ a_2 $ for which this occurs. The numerical scheme eventually fails at some radius, $r_{max}$ because the differential equation is very stiff. The point at which this failure occurs can be pushed to larger distance by choosing $a_2$ more precisely and increasing the working precision, albeit at the cost of increasing computation time\footnote{A solution for $r < r_{+}$ can be obtained by choosing $epsilon$ to be small and negative in Eq.(\ref{df}). The numerical scheme encounters no issues in this case.}.

\begin{figure}
	\begin{center}
		\includegraphics[height=4.45cm]{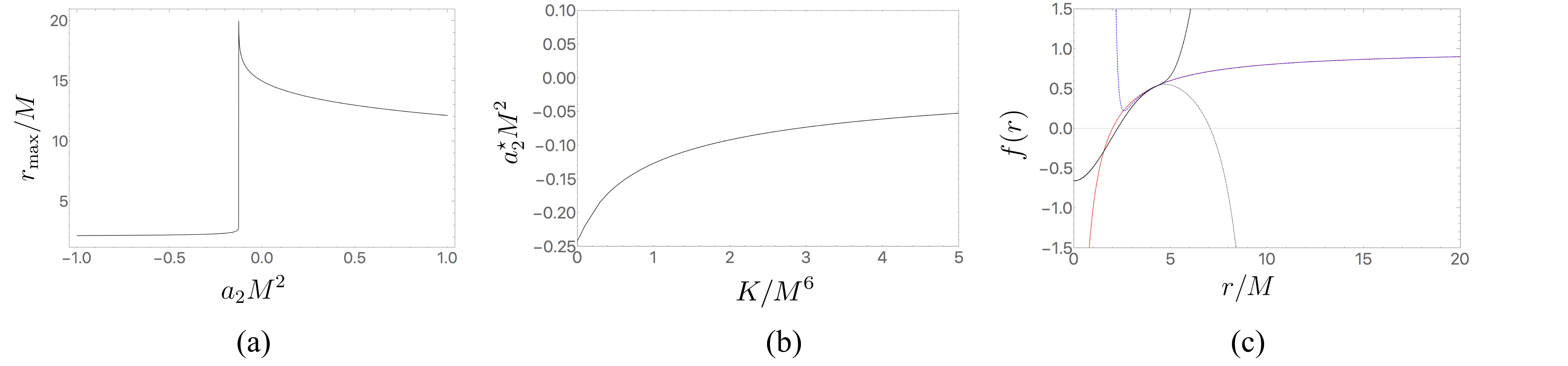}
		\caption{\small{\textbf{Numerical scheme:} \textbf{a:} A plot of $r_{max}$ (where the numerical solution breaks down) vs. $a_2$ for the case $K = 1$. The peak corresponds to the value of $ a_2 $ which gives an asymptotically flat solution. \textbf{b:} A plot of the value of $a_2$ gives an asymptotically flat solution vs. $K$. Note that in the limit $K\rightarrow0$ we have $a_2M^2\rightarrow -0.25$, which coincides with the Einstein gravity result. \textbf{c:} Numerical solution for $K/M^6 = 10$ and a shooting parameter $a^*_2 = -0.029981653451911665$. The solid, red curve is the Schwarzschild solution of Einstein gravity. The black, dotted curve is the near horizon approximation, including terms up to order $(r- r_{+})^8$. The dashed, blue curve is the asymptotic solution, including terms up to order $r^{-18}$. The solid black line is the numeric solution. In all cases, $\epsilon=10^{-6}$ was used in Eq.(\ref{df}) to obtain the initial data.}}
		\label{fig:0}
	\end{center}
\end{figure}

In figure~\ref{fig:0} we highlight some of our sample numerical results. Figure~\ref{fig:0}a depicts $r_{max}$ vs. $a_2$, showing a prominent peak at a point $a^{*}_2$. The peak coincides with the value of $a_2$ which produces the asymptotically flat solution. In  figure~\ref{fig:0}b, we plot $a^{*}_2$  against the coupling, $K$; as
expected,  when $K\rightarrow 0$, $a^{*}_2$ limits to the Schwarzschild value of $a^{*}_2M^2=-0.25$.  This behaviour is qualitatively similar to that found in ECG  \cite{Hennigar:2018hza}.
 We also get a fit of the numeric results as
\begin{align}
	a^{*}_2(x=K/M^6)=-\frac{1}{M^2}\frac{1 + 2.23817 x + 0.0322907 x^2}{4 + 15.0556 x + 6.70964 x^2},
\end{align}
which is accurate to three decimal places or better on the interval $K/M^6 \in [0, 5]$.  

In figure~\ref{fig:0}c we illustrate a comparison of our numerical solution for $K/M^6=10$ with  the near horizon and asymptotic approximate solutions along with  the  $K=0$ Schwarzschild solution (the red curve). For the same physical mass, the EQG black hole possesses a larger horizon radius than the Schwarzschild solution. Note the near horizon solution 
(the black dotted curve) gives an accurate approximation from $r = 0$ to about $r = 5M$, but rapidly diverges to $f \rightarrow -\infty$. Conversely, the asymptotic large-$r$ solution (the blue dashed curve) begins to break down near $r=3.5M$, but is otherwise fine at larger values of $r$; it is accurate to better than 1 part in 1,000 
and so   can be used to continue the solution to infinity. The numeric solution (the black solid curve) reproduces well the near horizon solution and 
 begins to rapidly converge to the asymptotic solution near $r = 4M$, but as $r \to 6M$ it breaks down: the stiff system causes the integrated solution to rapidly diverge to $f \rightarrow +\infty$ . This is just a result of not choosing $a_2$ to high enough precision in the numeric method, which ultimately excites the exponentially growing mode.  
 
Both the asymptotic and near horizon approximations have limitations, and the numeric solution is highly sensitive
to the choice of $a_2$.   Fortunately another approximation exists that yields an approximate solution valid everywhere outside of the horizon: the continued fraction approximation ~\cite{Rezzolla:2014mua, Kokkotas:2017zwt}.  To obtain it, we begin by changing coordinates 
\be 
x = 1 - \frac{r_+}{r},
\ee
so that the spacetime interval outside of the horizon is in the range $x \in [0,1)$. We then write
\be\label{eqn:cfrac_ansatz} 
f(x) = x \left[1 - \varepsilon(1-x) + (b_0 - \varepsilon)(1-x)^2 + \tilde{B}(x)(1-x)^3 \right],
\ee
where
\be 
\tilde{B}(x) = \cfrac{b_1}{1+\cfrac{b_2 x}{1+\cfrac{b_3 x}{1+\cdots}}} \, .
\ee
Inserting the ansatz~\eqref{eqn:cfrac_ansatz} in the field equations~\eqref{eqn:feq} yields
\begin{align}
\varepsilon &= \frac{2 M}{r_+} - 1 \qquad 
b_0 = 0 
\end{align}
at large $r$ $(x=1)$.  Furthermore,  expanding~\eqref{eqn:cfrac_ansatz} near the horizon ($x=0$), 
we find that all remaining coefficients are determined in terms of $T$, $M$, $r_+$ 
(consistent with \eqref{eqn:mass_equat} and \eqref{eqn:temp_equat}) 
and one free parameter, $b_2$.  Specifically
\begin{align}
b_1 &= 4 \pi r_+ T + \frac{4 M}{r_+} - 3 \qquad b_2 = - \frac{r_+^3 a_2 + 16 \pi r_+^2 T + 6(M-r_+)}{4 \pi r_+^2 T + 4 M - 3 r_+} 
\label{eqn:frac_b2}
\end{align}
and we see $b_2$ is given in terms of the coefficient $a_2$ appearing in the near horizon expansion~\eqref{eqn:nh_expand}.  All higher order coefficients are determined in terms of $T$, $M$, $r_+$ and $b_2$ (or, equivalently, $a_2$) from the field equations. Though their explicit form is quite cumbersome, they can easily be obtained using, e.g. \textrm{MAPLE}. The general expressions for the leading terms are given in the Appendix \ref{B}.
Note that we must manually input the value of  $b_2$ since it is not fixed by the field equations. This we do by  using the value of $a^{*}_2$ (as determined via the numerical method) in  (\ref{eqn:frac_b2}).

\begin{figure}
	\begin{center}
		\includegraphics[height=5cm]{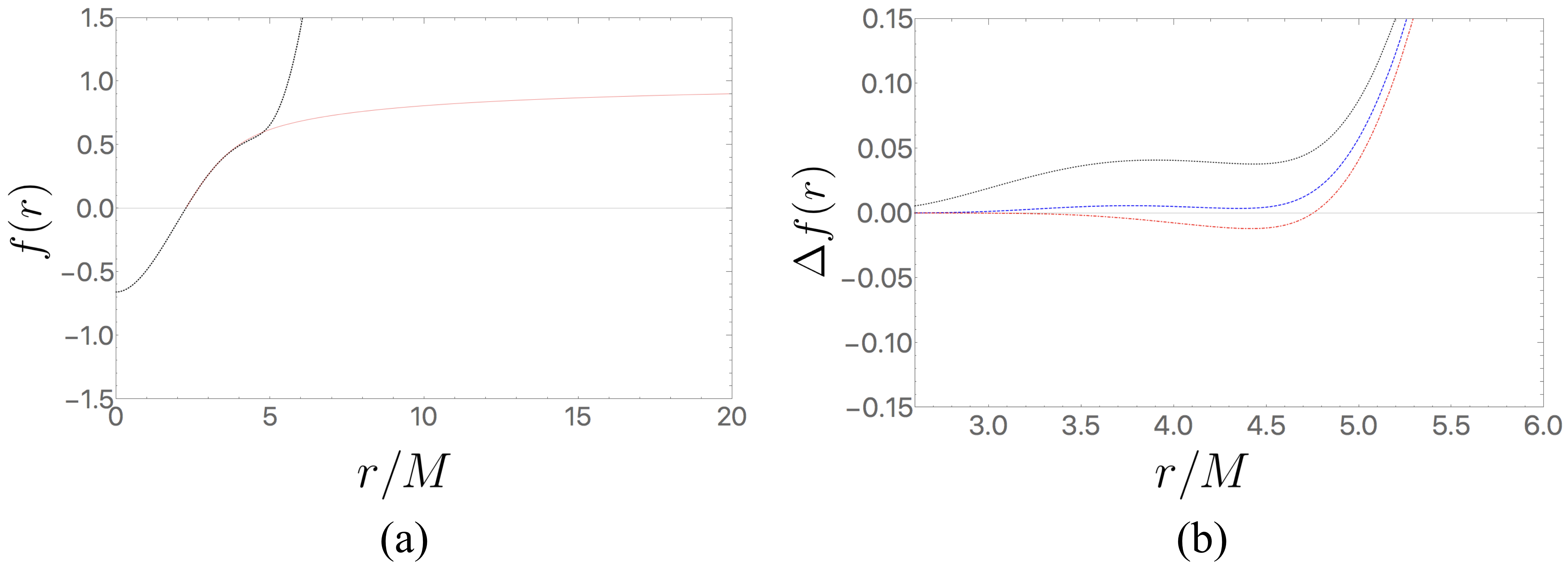}
		\caption{\small{\textbf{Continued fraction approximation: a:} Comparison of numeric solution (dotted, black) and continued fraction approximation (solid, red) for $K/M^6 = 10$. In the continued fraction, terms up to $b_5$ are kept; the continued fraction remains accurate even after the numeric solution fails. \textbf{b:} Difference between the metric function obtained numerically via the continued fraction approximation keeping terms up to $b_3$ (dotted, black), $b_4$ (dashed, blue), and $b_5$ (dot-dashed, red).}}
		\label{fig:1}
	\end{center}
\end{figure}

In contrast to   numerical integration of the field equations, which is highly sensitive to the precision with which $a^{*}_2$ is specified, the continued fraction provides a robust approximation  even with just a few digits of precision for $a^{*}_2$.  We illustrate the results in figure~\ref{fig:1}, where terms up to $b_5$ in the continued
fraction approximation have been retained.  The numeric solution breaks down at smaller values of $r/M$ than in  ECG  \cite{Hennigar:2018hza}, but the continued fraction accurately covers the entire region outside the horizon.
We show in figure~\ref{fig:1}b  the difference between the numerical solution and the continued fraction approximation -- where the numerical solution is valid, the continued fraction approximates it quite accurately.  

\section{PROPERTIES OF BLACK HOLE SOLUTIONS}
\label{sec:BH_props}

\begin{figure}
	\begin{center}
		\includegraphics[height=5cm]{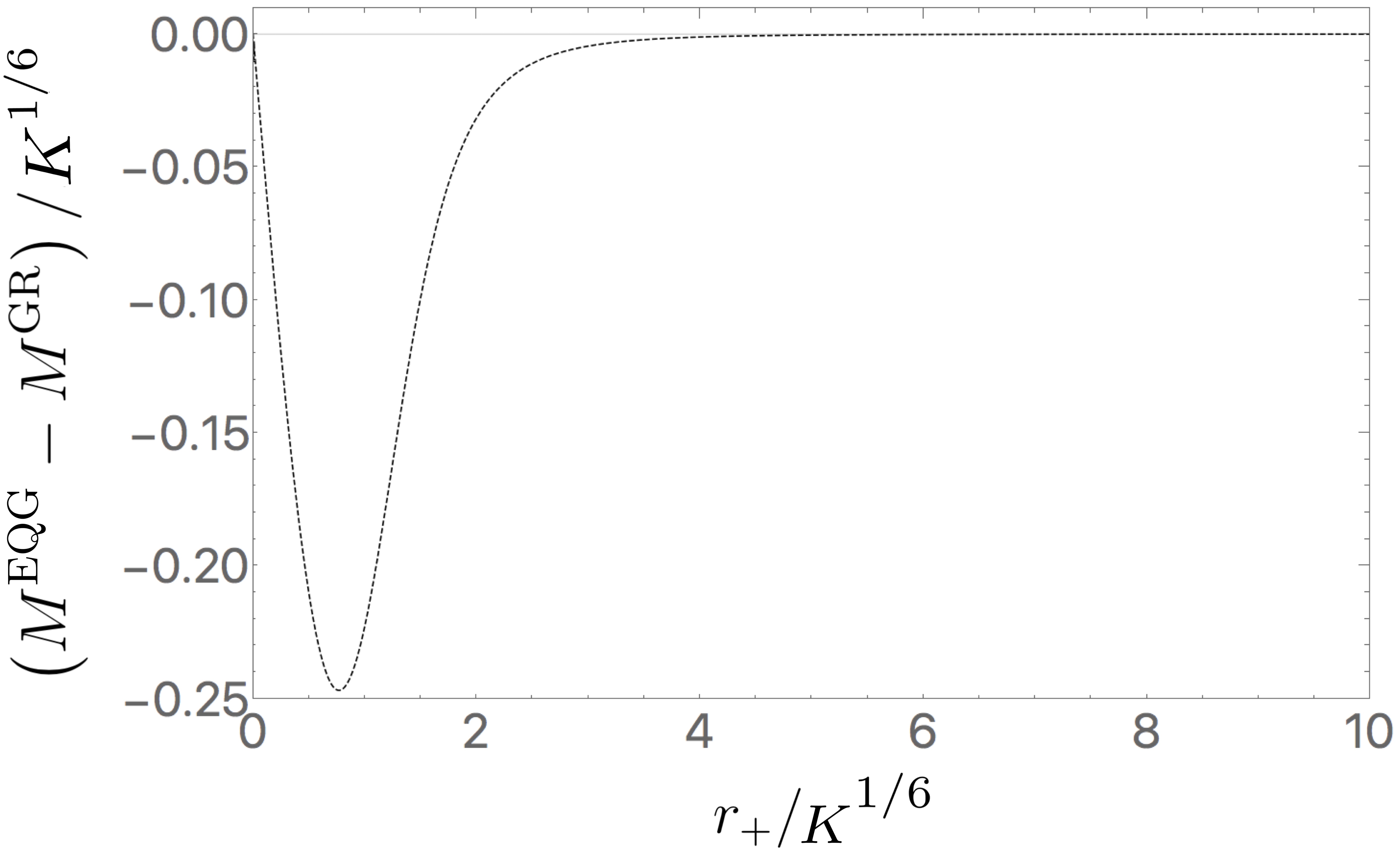}
		\caption{\small{\textbf{Mass vs. horizon radius:} a plot of the difference between the EQG and Einstein black hole mass vs. horizon radius. The plot shows a point of maximum difference.}}
		\label{fig:2}
	\end{center}
\end{figure}

In this section we consider some more interesting aspects of the black hole solutions in EQG. As in ECG, we have seen that although there is not an exact solution, the mass and temperature of these objects in EQG can be solved in terms of the horizon radius $r_+$ exactly from \eqref{eqn:mass_equat} and \eqref{eqn:temp_equat}. Furthermore, for fixed $K$, there is a particular value of $r_+$ for which the deviation
	from  Einstein gravity is largest. This is illustrated in figure~\ref{fig:2}. 
We find that this  maximal deviation occurs at $r^{dev}_+=(0.7695032663905271)K^{1/6}$. This in turn gives from
\eqref{eqn:mass_temp} a value of $M^{dev}=0.137868K^{1/6}$ for the mass of the black hole at this maximal deviation. The ratio $M^{dev}/r^{dev}_+$ is about $35.8\%$ that of the corresponding value in general relativity. 

\begin{figure}
	\begin{center}
		\includegraphics[height=4.5cm]{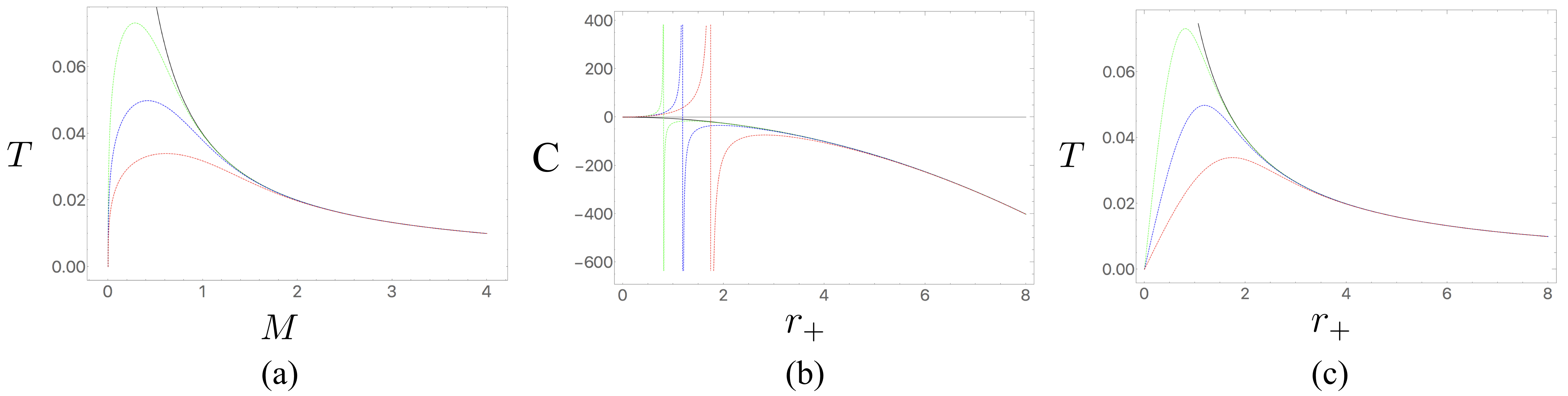}
		\caption{\small{\textbf{Thermodynamics of an EQG Black Hole} a: A plot of the the temperature vs. mass. b: A plot of the specific heat vs. the horizon radius. c: A plot of the the temperature vs. the horizon radius. In the all shapes, the dashed, red, blue and green curves are the EQG black holes in which coupling constant $K$ equals to $0.1$, $1$ and $10$ respectively. As well, the solid, black curves are the Schwarzschild solution of Einstein gravity.}}
		\label{fig:3}
	\end{center}
\end{figure}

The specific heat for sufficiently small mass for asymptotically flat ECG black holes is positive.
\cite{Bueno:2017qce,Hennigar:2018hza}.
EQG black holes share this feature, and in figure~\ref{fig:3}a we illustrate this by plotting 
the temperature of an EQG black hole as a function of its mass which its slope is the heat capacity
\begin{align}
C=\frac{\partial M}{\partial T} = 40 \pi r_+^2 + \frac{512 K \pi^3 T^2}{r_+^2} - \frac{
60 \pi r_+ (r_+^4 - 11 \pi r_+^5 T + 64 K \pi^2 T^2)}{
5 r_+^3 - 30 \pi r_+^4 T + 256 K \pi^3 T^3}\,,
\end{align}
using (\ref{eqn:mass_equat})  and (\ref{eqn:temp_equat}), and is
plotted in figure~\ref{fig:3}b against horizon radius. For a given $K$, this vanishes at a certain $r_+$ and $T_0$ which indicates that EQG black holes hotter than $T_0$ cannot exist which means above this temperature space is filled with pure radiation. Denoting by $r_0$ the value of $r_+$ where $T$ is maximal,
at any $T<T_0$ there are two black hole solutions: a stable smaller black hole, with $r<r_0$ 
having positive specific heat, and an unstable larger black hole with $r>r_0$, having  negative specific heat. 
In figure~\ref{fig:3}b we plot specific heat against the horizon radius $r_+$. For a given value of $K$ the specific heat diverges where
$T$ reaches its maximum $T_0$ at the critical point $r_0$, corresponding to a phase transition from a smaller EQG black hole with $C>0$ to a larger EQG black hole with $C<0$.  
Furthermore, from figure~\ref{fig:3}c, we see for any given $K > 0$ that temperature increases 
with increasing black hole radius up to the critical temperature $T_0$, yielding  $C>0$. 
For large values of $r_+$ the temperature decreases as the size of this black hole increases, and $C<0$.

Orbits of massive test bodies around an EQG black hole can be straightforwardly analyzed. Denoting by 
$\mu$ the rest mass of a test body, we have $g_{\alpha \beta}\dot{x}^\alpha \dot{x}^\beta = -\mu^2$, and
\begin{align}
\dot{r}^2 + f \left[1+\frac{\tilde{L}^2_z}{r^2}\right] =\tilde{E}^2\,,
\label{st}
\end{align}
from the geodesic equations, choosing coordinates so that the orbit lives on the equatorial plane, the overdot denoting the derivative with respect to proper time per unit rest mass \cite{mtw}. The quantities
  $\tilde{E}$ and $\tilde{L}$ are energy and angular momentum per unit rest mass $\mu$ of the body respectively. 

\begin{figure}
	\begin{center}
		\includegraphics[height=5cm]{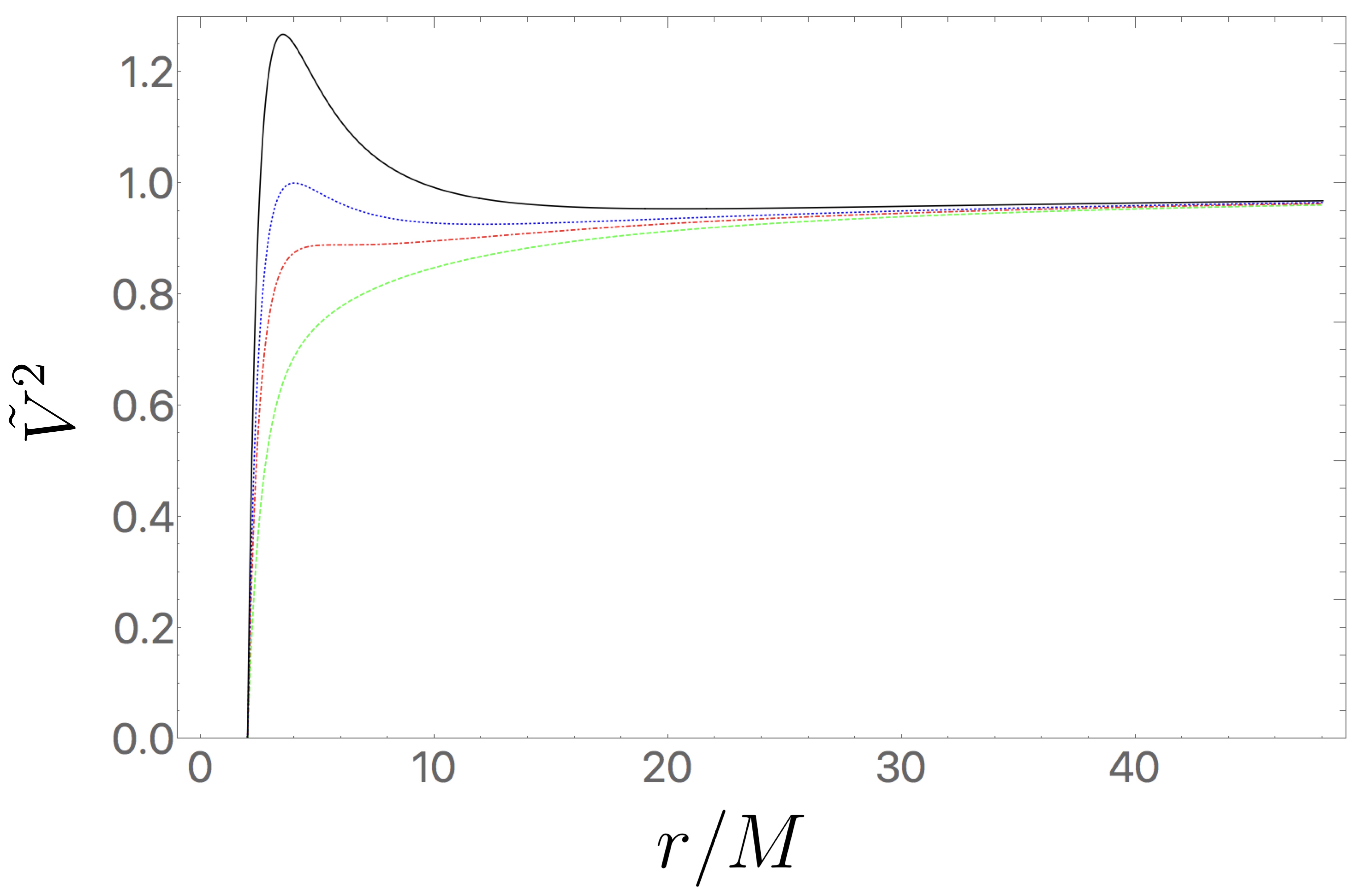}
		\caption{\small{\textbf{Effective potential of infalling particle:} For $K/M^6 = 0.1$ the effective potential is plotted for $\tilde{L}_{z,ISCO} \approx 2.4492M$ (green, dashed curve), $\tilde{L}_{z,ISCO} \approx 3.4638M$ (red, dot-dashed curve), $\tilde{L}_{z,ISCO} \approx 4.0002M$ (blue, dotted curve), and $\tilde{L}_{z,ISCO} \approx 4.8986M$ (black, solid curve). The blue dotted curve with $\tilde{L}_{z,ISCO} \approx 4.0002M$ has a maximum of 1. For a particle coming from infinity, $\tilde{L}_{z,ISCO} \approx 4.0002M$ is the minimum angular momentum it can have to avoid falling into the hole. The red, dot-dashed curve shows a point of inflection which is the innermost stable circular orbit.}
		}
		\label{fig:4}
	\end{center}
\end{figure}

 The second term  on the left hand side of equation (\ref{st}) 
\begin{align}
\tilde{V}^2=f\left[1+\frac{\tilde{L}^2_z}{r^2}\right],
\label{st}
\end{align}
is the effective gravitational potential experienced by the test body.
In figure~\ref{fig:4} we plot $\tilde{V}^2$ for $K/M^6 = 0.1$ for different values of $\tilde{L}_z$. For large values of $\tilde{L}_z$ there are two extrema in the curve of $\tilde{V}^2$, with the maximum (minimum) at the unstable (stable) orbits. By decreasing the value of $\tilde{L}_z$, the radius of the unstable equilibrium orbit increases and the radius of stable equilibrium orbit decreases. The ISCO is at the inflection point of $\tilde{V}^2$; this is $r = r_{\text\tiny{{ISCO}}}\approx 6.0029 $, which happens for particles with $\tilde{L}_z = \tilde{L}_{z,ISCO} \approx 3.4638M$. 
The numbers are numerically quite similar to  the ECG case.  This is due to the small value $K/M^6 = 0.1$ that was chosen.  When $K$ is sufficiently large the two theories substantively differ in their predictions.
The corresponding values in general relativity are $r_{ISCO} = 6M$ and $\tilde{L}_{z,ISCO}\approx 3.464M$. Recall that any bodies coming from infinity can be bounded only if $\tilde{V}^2 > 1$ (or equivalently if $\tilde{L}_{z,ISCO}\ge4.0002M$) since $ \tilde{E}\ge1 $. 

To find out how $r_{\tiny{ISCO}}$ and $\tilde{L}_{z,ISCO}$ change with $K$, we use a small $K$ approximation of the metric function
\begin{align}
f_{app}(r,K)=1-\frac{2M}{r}-\frac{1}{80}\frac{4661302698600944M^2-4917458785653298Mr+1300834826055069r^2}{r^3M^3(208200721714554M^2-217940342661245Mr+57257768532832r^2)} K
\label{small-k} 
\end{align}
The difference between this function and that obtained using the continued fraction up to $b_5$ is less than 1 part in 10,000 at $r_{\tiny{ISCO}}$ for $(K/M^6) < 1$. Noting that $r_{\tiny{ISCO}}$ is the inflection point of
\begin{align}
\tilde{V}_{app}^2=f_{app}\left(1+\frac{\tilde{L}_{z,ISCO}^2}{r^2} \right)
\label{infelection} 
\end{align}
we can obtain $r_{\tiny{ISCO}}$ and $\tilde{L}_{z,ISCO}$ for different values of $K$. 

\begin{figure}
	\begin{center}
		\includegraphics[height=5cm]{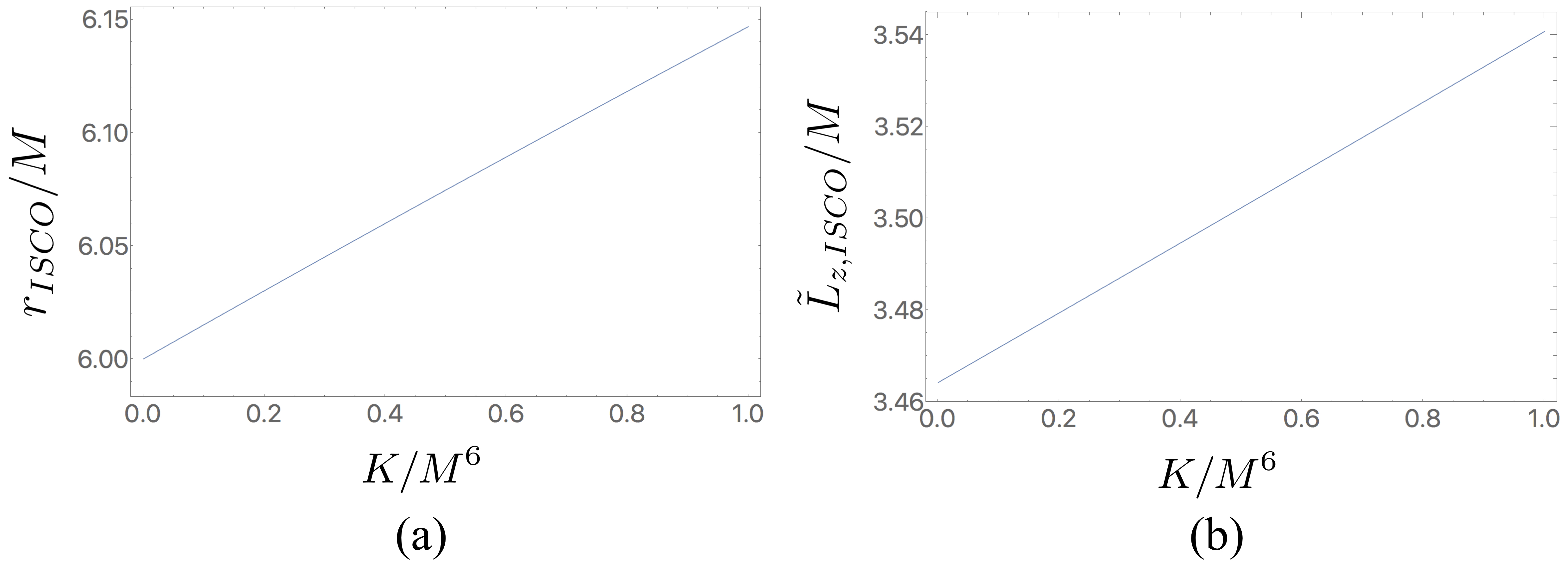}
		\caption{\small{\textbf{$r_{\tiny{ISCO}}/M$ and $\tilde{L}_{z,ISCO}/M$ vs. $K/M^6$}  a: Radius of the innermost stable circular orbit ($r_{\tiny{ISCO}}$) as a function of the coupling constant of EQG. b: The angular momentum for which the effective potential has an inflection point as a function of $K$.}
		}
		\label{fig:5}
	\end{center}
\end{figure}

In figure~\ref{fig:5}a we  plot $r_{\tiny{ISCO}}/M$ as a function of $K/M^6$. By fitting the numerical results we find the relation
\begin{align}
r_{\tiny{ISCO}}(K)/M\approx \frac{6+ 0.000359046 K/M^6}{1+ 0.0000345377 K/M^6}
\label{risco} 
\end{align}
 which is a small $K$ approximation of $r_{ISCO}$. Using (\ref{risco}) we obtain the approximate functional form 
\begin{align}
\frac{\tilde{L}_{z,ISCO}}{M}\approx\sqrt{\frac{12 + 0.29682754 K/M^6}{1 - 0.01911873 K/M^6}}
\label{Lisco} 
\end{align}
 of the angular momentum at the ISCO, shown in the bottom plot of figure~\ref{fig:5}b. We can see that by increasing $K$, both $r_{\tiny{ISCO}}/M$ and the angular momentum of the orbiting particle at $r=r_{\tiny{ISCO}}/M$ increase  relative to their values in general relativity. 
 
\section{CONSTRAINING EQG}
\label{sec:testEQG}

EQG corrections are most significant near the horizon, as the analysis in the previous sections makes clear. 
A generic black hole solution rapidly approaches its 
Schwarzschild counterpart  for distances a few times the horizon radius. Since (as in Einstein gravity) the post-Newtonian parameter $\gamma$ is unity in EQG,  deviations imposed by EQG will be small   in the weak-field regime. 

Consider first how Shapiro time delay, the most accurate of the Solar System tests, constrains EQG.  The time for a photon to travel between the points $r_0$ and $r$ is given by the integral \cite{weinberg1972}
\begin{align}
t(r,r_0)= \int_{r_0}^r \frac{dr/f(r)}{\sqrt{ 1-(r_0/r)^2 (f(r)/f(r_0))}}
\label{eqn:delay0} 
\end{align}
where it is straightforward to do the integration numerically and this is the method we use.

However  for all practical purposes, the first few terms in the asymptotic expansion in section \ref{seclgr} 
for $f(r)$ can be used. It is illuminating to consider an analytic approximation to \eqref{eqn:delay0}, which takes the form  
\begin{align}
t(r,r_0)=t^{\text{SR}}(r,r_0)+\Delta t^{\text{GR}}(r,r_0)+\Delta t^{\text{EQG}}(r,r_0) 
\label{eqn:delay1},
\end{align}
 where the special relativistic contribution $t^{\text{SR}}(r,r_0)=\sqrt{r^2-r_0^2}$ comes from light propagating in flat space-time. The general relativistic correction to this 
\begin{align}
\Delta t^{\text{GR}}(r,r_0)=2M \ln{\left[\frac{r+\sqrt{r^2-r_0^2}}{r_0}\right]}+M\sqrt{\frac{r-r_0}{r+r_0}}+...
\label{eqn:delay2},
\end{align}
is well known, with higher-order corrections straightforwardly computed.  The EQG correction is 
\begin{align}
\Delta t^{\text{EQG}}(r,r_0)=\frac{K}{M^6}\left[\frac{432 M^9\left(3 r_0^8+2 r_{0}^6 r^2+4 r_0^4 r^4+16 r_0^2 r^6+7 r_0 r^7-32 r^8\right)}{35 r^7 r_0^8\sqrt{(r^2-r_0^2)}}\right],
\label{eqn:delay3}
\end{align}
to leading order in  $M/r$, $M/r_0$ and $K/M^6$.

We see from \eqref{eqn:delay3} how enormously suppressed EQG corrections   are at the level of Solar System tests.  Taking $M$ to be a solar mass ($M_\odot=1477$m) and choosing  $r_0$ to be the radius of the Sun, $r_\odot=6.957\times10^8$ m, the factor in square brackets is on the order of  $100 M_\odot\left(M_\odot/r_\odot \right)^8\sim 10^{-44} M_\odot.$   So $K/M^6$ can be very large while being consistent with Solar System tests of general relativity.

For a radar signal traveling from Earth to Mercury, grazing the Sun along the way, the time delay is 
\begin{align}
(\Delta t)_{\text{max}}=2\left[ t(r_{\earth},r_\odot)+t(r_\odot,r_{\mercury})-\sqrt{r_{\earth}^2-r_\odot^2} -\sqrt{r_{\mercury}^2-r_\odot^2}\right],
\end{align}
whose deviation from general relativity is empirically less than $0.0012\%$ \cite{Will:2014kxa}. A precise numerical evaluation of the integrals determines that if  
\begin{align}
K < 8.98 \times 10^{38} M_\odot^6
\end{align}
then EQG will be consistent with constraints coming from the Shapiro time delay experiment. The astonishing size of this value clarifies that the deviations from general relativity are most important in the vicinity of a black hole horizon. 


\section{BLACK HOLE SHADOWS and OBSERVATIONAL TESTS OF EQG}
\label{sec:shadow}

Since the continued fraction approximate solution is valid everywhere outside the horizon  it can be employed just
as though it were an exact analytic solution. We shall construct in this section an equation for the angular radius of the black hole shadow as seen by a distant observer, something that would be considerably more difficult were we to 
use  only the numerical solution.  

For the spherically symmetric line element in Eq.~\eqref{metsss} the Lagrangian  is
\begin{align}\label{Lagnul}
\mathcal{L}=\frac{1}{2}g_{\mu \nu} \dot{x}^\mu \dot{x}^\nu=\frac{1}{2} (-f \dot{t}^2+\frac{\dot{r}^2}{f}+r^2\sin^2 \theta \dot{\phi}^2) 
\end{align}
yielding from the equations of motion
\begin{align}
E=-\frac{\partial \mathcal{L}}{\partial \dot{t}}=f\dot{t}, \hspace{2cm} L_z=-\frac{\partial \mathcal{L}}{\partial \dot{\phi}}=r^2\dot{\phi}^2,
\end{align}
as the respective conserved  energy and angular momentum 
of a light ray travelling toward the black hole. Without loss of generality, we have chosen coordinates so that the light ray is in the equatorial plane.  

 For null geodesics $\mathcal{L}=0$, after some calculations we can write \eqref{Lagnul}   as \cite{synge} 
\begin{align}\label{s1}
\left(\frac{dr}{d\phi}\right)^2=r^4\left(\frac{1}{\xi^2}-\frac{f}{r^2}\right),
\end{align}
 for $\theta=\pi/2$, using $\xi=L_z/E$ as a constant of the motion.  The shadow of the black hole
 (or alternatively the photon sphere)  is at the radius $r_*$ where $\frac{r^2}{f}$ is minimized, ensuring
 that if $\xi^2$ is less than $\frac{r_*^2}{f(r_*)}$ the light ray always reaches the horizon since its 
 $r$ coordinate  always decreases.
 Denoting the inclination angle of the light ray from the radial direction by $\delta$, we write  \cite{synge} 
\begin{align}\label{s2}
\cot \delta=\frac{1}{\sqrt{f}r} \left(\frac{dr}{d\phi}\right)
\end{align}
and from  (\ref{s1}) and (\ref{s2}) we obtain
 \begin{align}\label{s3}
\delta=\sin^{-1}\left(\sqrt{\frac{r_{ps^2}}{f(r_{ps})}\frac{f(D)}{D^2}}\right)
\end{align}
for the angular radius of the shadow as seen by an observer located at $D$, where the radius of the photon sphere is $r_{ps}$.

The current EHT project \cite{ehtt}  will study the black hole at the center of our Galaxy. Present-day observation represents that this black hole, Sgr A*, has a mass $M =6.25\times 10^9$ m and its distance is $D=2.57\times10^{20}$ m \cite{mnd}. In general relativity its horizon radius is $r_+ = 2M$, and the radius of its photon sphere is $r_{ps} = 3M$. Using (\ref{s3}) we obtain the known result $\delta = 26.05$ $\mu$as. 

It is clear that  for sufficiently small $K$ it will pass all solar system tests. However for large enough black hole masses
(such as Sgr A*) its predictions will depart from those of general relativity.
\begin{figure}
	\begin{center}
		\includegraphics[height=5cm]{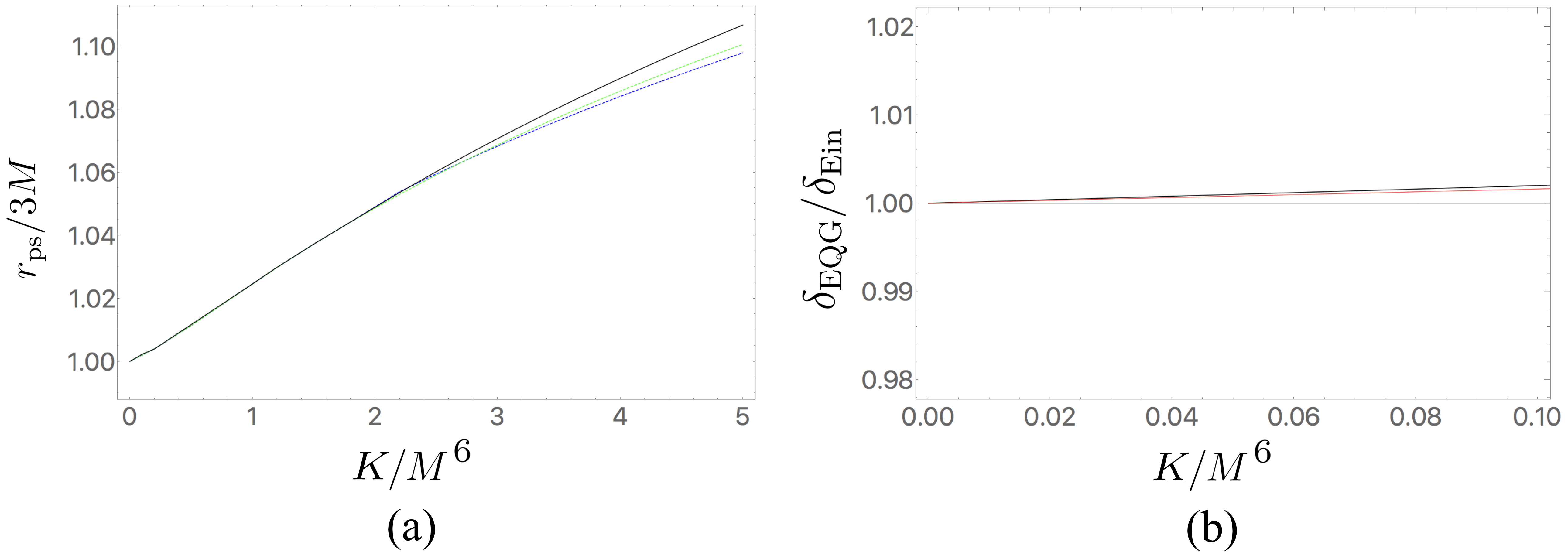}
		\caption{\small{\textbf{Photon sphere and angular radius} a: A plot of the photon sphere radius, $r_{ps}$, vs. the EQG coupling computed using the continued fraction truncated at: b2 (blue,dotted), b4 (green, dashed) and b6 (black, solid). For small coupling (compared to the mass), even the lowest order approximation is accurate, while for larger couplings the continued fraction converges after the inclusion of the first few terms (the black and green curves are virtually indistinguishable). The general result is that larger EQG coupling pushes the photon sphere to larger distances. b: A plot of the ratio of the angular radius of the shadow for a black hole of mass $6.25 \times 10^9$ m and viewing distance $D = 2.57 \times 10^{20}$ m. The solid black line is the result of the continued fraction, truncated with $b_5 = 0$. The red line is a linear approximation for small coupling, shown in Eq. (\ref{angular}).}}
		\label{fig:6}
	\end{center}
\end{figure}

If EQG is correct, then the   metric outside of a spherically symmetric black hole will be given by~\eqref{metsss} with the 
metric function $f$ approximated to excellent accuracy  by the continued fraction approximation (\ref{eqn:cfrac_ansatz}).
The quantities  $M$ and $D$ will have the values given above for  Sgr A*, and  
equation (\ref{eqn:mass_temp}) indicates that the horizon radius in EQG will be larger than in general relativity  and will be smaller than in ECG.  The  radius of the photon sphere in EQG is likewise larger than in general relativity as shown in figure~\ref{fig:6}a, but it is smaller than in ECG \cite{Hennigar:2018hza}. 

It is a simple matter to compute the angular radius of the black hole shadow using (\ref{s3}). We present results of this calculation, as specified through a continued fraction approximation truncated at $b_5$ in figure~\ref{fig:6}b for choices of mass and distance relevant for Sgr A*. It should be noted that, when $K/M^6$ is small, we can use Eq. (\ref{small-k}) to obtain the following expansion for the angular radius of the shadow
\begin{align}\label{angular}
\delta_{EQG}=\delta_{Ein}+\frac{2.98975\times10^{10}}{\sqrt{\frac{3D^2}{f(D) M^2}-81}} \frac{K}{M^6}+\mathcal{O}(K^2)
\end{align}
 (in $\mu$as) which appears in figure~\ref{fig:6}b as the red curve and is different less than one percent from the numerical results for any $K\leq 0.1$. EQG predicts larger black hole shadows than Einstein gravity. Since, for larger distances, $f(D)$ is practically identical in both general relativity and EQG, the differences in figure~\ref{fig:6}b are the modifications result in the strong gravity regime near the horizon. However, as expected from dimensional grounds, for objects of large mass, the modifications are relatively small, which needs $K/M^6 \approx 0.5$ before occuring differences of $1\%$.

\section{Conclusions}
\label{sec:con}

We have carried out the first phenomenological study of Einstein Quartic Gravity, a class of theories whose corrections to the Einstein-Hilbert action  are quartic in the curvature. Under spherical symmetry there is a degeneracy that yields the same field equation for all theories in the class, meaning that the corrections to Einstein gravity depend on a single coupling parameter. These theories are the next simplest kinds that occur in Generalized Quasi-Topological Gravity, and are the highest in curvature that allow explicit solutions for the mass and temperature in terms of the horizon radius, as shown in \eqref{eqn:mass_temp}.  
 
We have obtained solutions in both the near-horizon and large distance approximations, as well as numerically using the shooting method.  We have also obtained a  continued fraction approximation \eqref{eqn:cfrac_ansatz}, and have shown that this approximation    accurately describes  black hole solutions in EQG everywhere outside the horizon. The  continued fraction approximation has a distinct advantage insofar as it allows approximate analytic treatment
of various scenarios (such as geodesic motion, shadows, etc.) everywhere outside the horizon.
 
Phenomenologically we find that for small values of coupling constant, EQG black holes have considerable resemblance to their ECG counterparts 
\cite{Hennigar:2018hza}.
For a given value of mass, the ISCO for a massive test body is on a larger radius for a larger value of the coupling constant $K$ in EQG. Likewise, the angular momentum of the body at the ISCO increases with increasing $K$ and study of the lightlike geodesics shows that EQG enlarges the shadow of the black hole  relative to Einstein gravity, whereas the shadow in EQG is smaller than in ECG.  

\begin{figure}
	\begin{center}
		\includegraphics[height=5cm]{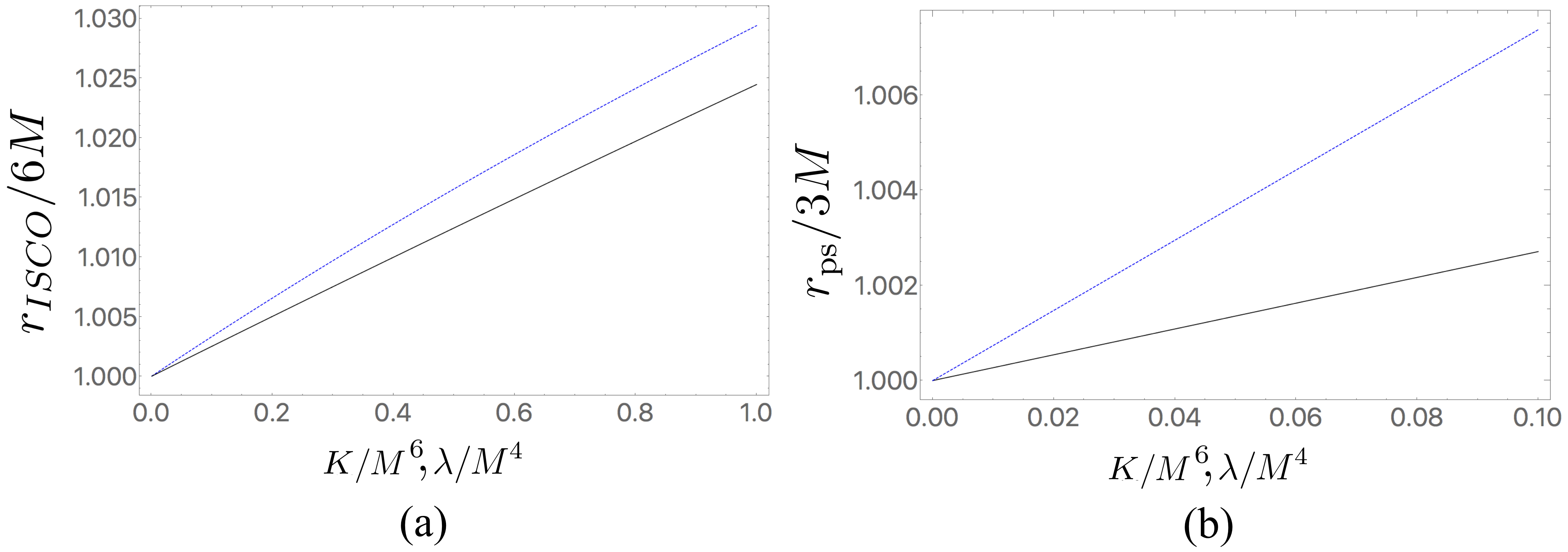}
		\caption{\small{\textbf{Comparison between EQG and ECG} a: Radius of the innermost stable circular orbit ($r_{ISCO}$) as a function of the dimensionless (small) coupling constants in EQG (solid black line) and ECG (dotted blue line). b: The photon sphere radius ($r_{ps}$) as a function of small coupling constant compared to the mass in EQG (solid black line) and ECG (dotted blue line).}}
		\label{fig:7}
	\end{center}
\end{figure}

In figure~\ref{fig:7} we compare EQG, ECG and general relativity   for the  quantities $r_{ISCO}$ and $r_{ps}$ for small values of the respective coupling constants. We see that $r_{ISCO}$ and $r_{ps}$ in EQG are slightly smaller than in ECG and they increase as coupling constant increases.  This indicates that observations of sufficient precision to
detect ECG effects can likewise detect EQG effects and furthermore distinguish between them.

As expected from dimensional analysis, the effect of EQG is relatively small unless the dimensionless ratio $K/M^6$ becomes significantly large, as indicated in our study of Shapiro time delay. We therefore expect the effects of EQG to be most significant for strong gravitational fields, and so considered the structure of black hole shadows.   
 For Sgr A* we find that for the largest value of $K$ allowed by Shapiro time delay, EQG enlarges the angular radius of the shadow by about 2 parts per thousand. This is more than the angular radius of the shadow in ECG because the largest value of the dimensionless coupling constant $K/M^6$ allowed by Shapiro time delay in EQG is more than its counterpart in ECG \cite{Hennigar:2018hza}. 
  Generally for a same value of coupling constant, the angular radius of the shadow in EQG is smaller than in ECG.  Nowadays, the resolution of EHT’s 1.3 mm groundbased very long baseline interferometry (VLBI) is a few tens of microarcseconds \cite{Akiyama} which is about the shadow size of Sgr A* and M87. In this resolution the shadow predicted by general relativity and EQG are indistinguishable at least for static solutions. 
By increasing the maximum distance in a VLBI array, i.e. by adding some space stations, or observing at shorter wavelengths, EHT (or a similar project) might achieve finer resolutions in the future. However resolutions of better than 10 nanoarcseconds to observe the effects of EQG on Sgr A* shadow will be required.

A natural direction for future work would involve extending these results to compute shadows of rotating black holes in EQG, analogous to what was recently done in ECG \cite{Adair:2020vso}. Rotating black holes are of more direct astrophysical relevance, and may present distinct angular-dependent features that could be observed.  For sufficiently slow rotation, there is an additional metric function in the   $g_{t\phi}$ component in   (\ref{metsss}).
For ECG there were   two second-order differential field equations for the two metric functions that were be solved using the methods presented here. We expect a similar situation to hold for EQG, but with the degeneracy  in \eqref{EQGK} broken. This will allow us to see how different quartic theories can be empirically distinguished. Similar techniques as those presented here (see also \cite{Younsi:2016azx}) could be used to obtain an approximate rotating black hole solutions in this theory. 

Another possibility would be to consider higher-curvature theories.  One explicit example of a GQTG that is quintic in curvature has been constructed~\cite{Bueno:2019ycr}, though the most general GQTG action quintic in curvature is not known.  Fortunately it is not necessary to explicitly construct all possible actions since a general expression for the field equation for a spherically symmetric black hole in any dimension for an $n$-th order GQTG is known~\cite{Bueno:2019ycr}.  It would be interesting to carry out a study of the type we have carried out here for this general case to see what features emerge.

 The continued fraction approach also offers the exciting possibility of addressing the linear stability of black hole solutions in this theory by simplifying the analysis of quasi-normal modes. Not only would this be of astrophysical relevance for the four dimensional models, but it would also be relevant in the context of holography for the asymptotically AdS solutions. We hope to address these and other questions in future work. 

\section*{Acknowledgements}
This work was supported in part by the Natural Sciences and Engineering Research Council of Canada.  We are grateful to Robie Hennigar for helpful discussion and correspondence.

\onecolumngrid

\appendix
\section{Generalized quasi-topological Lagrangian densities}\label{A}

Here we present the explicit forms of the six terms that we added to the Einstein-Hilbert Lagrangian density~\eqref{actionEQG}.
\begin{align}
\mathcal{S}_4^{(1)} &= R_{\mu}\,^{\xi}\,_{\rho}\,^{\upsilon} R^{\mu \nu \rho \sigma} R_{\nu}\,^{\omega}\,_{\xi}\,^{\tau} R_{\sigma \omega \upsilon \tau} + \frac{1}{1080} R_{\mu}\,^{\rho} R^{\mu \nu} R_{\nu}\,^{\sigma} R_{\rho \sigma} + \frac{16}{5} (R_{\mu \nu} R^{\mu \nu})^2 
\nn\\ 
&-\frac{1}{90} R_{\mu}\,^{\rho} R^{\mu \nu} R_{\nu \sigma} R - \frac{37}{45} R^2 R_{\mu \nu} R^{\mu \nu} + \frac{1}{18} R^2 R_{\mu \nu \rho \sigma} R^{\mu \nu \rho \sigma} + \frac{86}{45} R R^{\mu \nu} R^{\rho \sigma} R_{\mu \rho \nu \sigma} 
\nn\\ 
&-\frac{208}{15} R_{\mu}\,^{\rho} R^{\mu \nu} R^{\sigma \xi} R_{\nu \sigma \rho \xi}+\frac{82}{15} R^{\mu \nu} R^{\rho \sigma} R_{\mu \rho}\,^{\sigma \upsilon} R_{\nu \sigma \xi \upsilon} - \frac{1}{360} R R_{\mu \nu}\,^{\xi \upsilon} R^{\mu \nu \rho \sigma} R_{\rho \sigma \xi \upsilon} 
\nn\\
&-\frac{8}{9} R_{\mu \nu} R^{\mu \nu} R_{\rho \sigma \xi \upsilon} R^{\rho \sigma \xi \upsilon} -\frac{1}{3} R^{\mu \nu} R_{\mu}\,^{\rho \sigma \xi}  R_{\nu \rho}\,^{\upsilon \omega} R_{\sigma \xi \upsilon \omega} \,, 
\end{align}
\begin{align}
\mathcal{S}_4^{(2)} &= R_{\mu}\,^{\xi}\,_{\rho}\,^{\upsilon} R^{\mu \nu \rho \sigma} R_{\nu}\,^{\omega}\,_{\sigma}\,^{\tau} R_{\xi \omega \upsilon \tau} + 2 R_{\mu}\,^{\rho} R^{\mu \nu} R_{\nu}\,^{\sigma} R_{\rho \sigma} + \frac{5}{2} (R_{\mu \nu} R^{\mu \nu})^2
\nn\\
&-\frac{1}{45} R_{\mu}\,^{\rho} R^{\mu \nu} R_{\nu \rho} R - \frac{3}{4} R^2 R_{\mu \nu} R^{\mu \nu} + \frac{1}{8} R^2 R_{\mu \nu \rho \sigma} R^{\mu \nu \rho \sigma} + 4 R R^{\mu \nu} R^{\rho \sigma} R_{\mu \rho \nu \sigma} 
\nn\\
&-14 R_{\mu}\,^{\rho} R^{\mu \nu} R^{\sigma \xi} R_{\nu \sigma \rho \xi}+5 R^{\mu \nu} R^{\rho \sigma} R_{\mu \rho}\,^{\xi \upsilon} R_{\nu \sigma \xi \upsilon} - \frac{1}{4} R R_{\mu \nu}\,^{\xi \upsilon} R^{\mu \nu \rho \sigma} R_{\rho \sigma \xi \upsilon}
\nn\\
&-\frac{3}{4} R_{\mu \nu} R^{\mu \nu} R_{\rho \sigma \xi \upsilon} R^{\rho \sigma \xi \upsilon} - R^{\mu \nu} R_{\mu}\,^{\rho \sigma \xi}  R_{\nu \rho}\,^{\upsilon \omega} R_{\sigma \xi \upsilon \omega} \,,
\end{align}
\begin{align}
\mathcal{S}_4^{(3)} &= R_{\mu \nu}\,^{\xi \upsilon} R^{\mu \nu \rho \sigma} R_{\rho \xi}\,^{\omega \tau} R_{\sigma \upsilon \omega \tau} -\frac{6}{5} R_{\mu}\,^{\rho} R^{\mu \nu} R_{\nu}\,^{\sigma} R_{\rho \sigma} + \frac{1}{5} (R_{\mu \nu} R^{\mu \nu})^2
\nn\\ 
&+\frac{6}{5} R_{\mu}\,^{\rho} R^{\mu \nu} R_{\nu \rho} R - \frac{1}{10} R^2 R_{\mu \nu} R^{\mu \nu}  -\frac{1}{4} R^2 R_{\mu \nu \rho \sigma} R^{\mu \nu \rho \sigma} + \frac{4}{5} R R^{\mu \nu} R^{\rho \sigma} R_{\mu \rho \nu \sigma}
\nn\\
&-\frac{36}{5} R_{\mu}\,^{\rho} R^{\mu \nu} R^{\sigma \xi} R_{\nu \sigma \rho \xi}+\frac{24}{5} R^{\mu \nu} R^{\rho \sigma} R_{\mu \rho}\,^{\xi \upsilon} R_{\nu \sigma \xi \upsilon} + \frac{1}{10} R R_{\mu \nu}\,^{\xi \upsilon} R^{\mu \nu \rho \sigma} R_{\rho \sigma \xi \upsilon}
\nn\\
&+\frac{1}{2} R_{\mu \nu} R^{\mu \nu} R_{\rho \sigma \xi \upsilon} R^{\rho \sigma \xi \upsilon}
- 2R^{\mu \nu} R_{\mu}\,^{\rho \sigma \xi}  R_{\nu \rho}\,^{\upsilon \omega} R_{\sigma \xi \upsilon \omega} \,,
\end{align} 
\begin{align}
\mathcal{S}_4^{(4)} &= R_{\mu \nu}\,^{\xi \upsilon} R^{\mu \nu \rho \sigma} R_{\rho \sigma}\,^{\omega \tau} R_{\xi \upsilon \omega \tau} -\frac{12}{5} R_{\mu}\,^{\rho} R^{\mu \nu} R_{\nu}\,^{\sigma} R_{\rho \sigma} + \frac{2}{5}(R_{\mu \nu} R^{\mu \nu})^2  
\nn\\
&+\frac{12}{5} R_{\mu}\,^{\rho} R^{\mu \nu} R_{\nu \rho} R - \frac{1}{5} R^2 R_{\mu \nu} R^{\mu \nu}  -\frac{1}{2} R^2 R_{\mu \nu \rho \sigma} R^{\mu \nu \rho \sigma} + \frac{8}{5} R R^{\mu \nu} R^{\rho \sigma} R_{\mu \rho \nu \sigma}
\nn\\
&-\frac{72}{5} R_{\mu}\,^{\rho} R^{\mu \nu} R^{\sigma \xi} R_{\nu \sigma \rho \xi}+\frac{48}{5} R^{\mu \nu} R^{\rho \sigma} R_{\mu \rho}\,^{\xi \upsilon} R_{\nu \sigma \xi \upsilon} + \frac{1}{5} R R_{\mu \nu}\,^{\xi \upsilon} R^{\mu \nu \rho \sigma} R_{\rho \sigma \xi \upsilon}
\nn\\
&+R_{\mu \nu} R^{\mu \nu} R_{\rho \sigma \xi \upsilon} R^{\rho \sigma \xi \upsilon}
- 4R^{\mu \nu} R_{\mu}\,^{\rho \sigma \xi}  R_{\nu \rho}\,^{\upsilon \omega} R_{\sigma \xi \upsilon \omega} \,,
\end{align}
\begin{align}
\mathcal{S}_4^{(5)} &= -\frac{14}{5} (R_{\mu \nu} R^{\mu \nu})^2 -\frac{20}{3} R_{\mu}\,^{\nu} R_{\nu}\,^{\rho} R_{\rho}\,^{\sigma} R_{\sigma}\,^{\mu} - \frac{8}{5} R R^{\mu \rho} R^{\nu \sigma} R_{\mu \nu \rho \sigma} 
\nn\\
&+\frac{104}{5} R^{\mu \nu} R_{\xi}\,^{\sigma} R^{\xi \rho} R_{\mu \rho \nu \sigma}+R_{\xi \upsilon} R^{\xi \upsilon} R_{\mu \nu \rho \sigma} R^{\mu \nu \rho \sigma} + \frac{1}{5} R^2 R^{\mu \nu \rho \sigma} R_{\mu \nu \rho \sigma} 
\nn\\
&-\frac{56}{15} R^{\mu \nu} R_{\rho \sigma}\,^{\omega}\,_{\mu} R^{\rho \sigma \xi \upsilon} R_{\xi \upsilon \omega \nu} + R_{\mu \nu \rho}\,^{\xi} R^{\mu \nu \rho \sigma} R_{\upsilon \omega \tau \sigma} R^{\upsilon \omega \tau}\,_{\xi} \,,
\end{align}
\begin{align}
\mathcal{S}_4^{(6)} &= -\frac{308}{15} (R_{\mu \nu} R^{\mu \nu})^2 -\frac{64}{3} R_{\mu}\,^{\nu} R_{\nu}\,^{\rho} R_{\rho}\,^{\sigma} R_{\sigma}\,^{\mu} + \frac{64}{15} R R^{\mu \rho} R^{\nu \sigma} R_{\mu \nu \rho \sigma}
\nn\\
&+\frac{1088}{15} R^{\mu \nu} R_{\xi}\,^{\sigma} R^{\xi \rho} R_{\mu \rho \nu \sigma} + \frac{28}{3}R_{\xi \upsilon} R^{\xi \upsilon} R_{\mu \nu \rho \sigma} R^{\mu \nu \rho \sigma} - \frac{8}{15} R^2 R^{\mu \nu \rho \sigma} R_{\mu \nu \rho \sigma} 
\nn\\
&-\frac{224}{15} R^{\mu \nu} R_{\rho \sigma}\,^{\omega}\,_{\mu} R^{\rho \sigma \xi \upsilon} R_{\xi \upsilon \omega \nu} + (R_{\mu \nu \rho \sigma} R^{\mu \nu \rho \sigma})^2 \,.
\end{align} 

\section{Explicit Terms in Continued Fraction}\label{B}

Here we present additional terms that appear in the continued fraction expansion~\eqref{eqn:cfrac_ansatz}.
\begin{align}
b_3= - &\frac{1}{9216} \frac{1}{K (\pi r_+ T +\frac{1}{2})(r_+^2 T \pi+M -\frac{3}{4}r_+ )T^2r_+^2 \pi^2 b_2}\Bigl[ 20\pi T (b_2+3) r_+^8+(-15b_2 -30)r_+^7
\nn\\
&+(20 M b_2+30M)r_+^6+21504K \Bigl(b_2^2+\frac{160}{21}b_2+\frac{487}{28} \Bigr) T^4 \pi^4 r_+^5-17664 K T^3 \Bigl(b_2^2 +\frac{136}{23}b_2
\nn\\
&+\frac{122}{23}\Bigr)\pi^3 r_+^4+33792 K  T^2 \Bigl(M \Bigl(b_2^2+\frac{196}{33} b_2+\frac{74}{11} \Bigr) T\pi-\frac{15}{22}b_2-\frac{57}{44}-\frac{3}{88}b_2^2 \Bigr) \pi^2 r_+^3
\nn\\
&-7680 K  T  \Bigl(M T \Bigl (b_2^2-\frac{12}{5} \Bigr) \pi -\frac{9}{40}(b_2+2)^2 \Bigr)\pi r_+^2
\nn\\
&-12288\Bigl(-b_2-\frac{3}{2}\Bigr) M K \Bigl(\Bigl (b_2+\frac{3}{2}\Bigr) MT \pi-\frac{3}{8} b_2-\frac{3}{4}\Bigr) T \pi r_+ 
\nn\\
&+3072\Bigl(b_2+\frac{3}{2}\Bigr)^2 M^2 K  T \pi \Bigr] \,,
\end{align}
 
\begin{align}
b_4 &= -\,{\frac {1}{24576  \Bigl( \pi \,r_+T+\frac{1}{2} \Bigr){K} b_{3} \Bigl({r_+}^{2}T\pi +M -\frac{3}{4}\,r_+ \Bigr){T}^2{r_+}^4{\pi}^2 b_{2}}}
\nn\\
&\times \Bigl[ 20\,T \Bigl( 6+{b_{{2}}}^{2}+ \Bigl( b_{{3}}+4 \Bigr) b_{{2}} \Bigr) 
\mbox{}\pi \,{r_+}^{10}+ \Bigl( -70-15\,{b_{{2}}}^{2}+ \Bigl( -15\,b_{{3}}-60 \Bigr) b_{{2}} \Bigr) {r_+}^{9}
\nn\\
&\mbox{}+20\,M \Bigl( 4+{b_{{2}}}^{2}+ \Bigl( b_{{3}}+4 \Bigr) b_{{2}} \Bigr) 
\mbox{}{r_+}^{8}+129024\,K \Bigl( {\frac {1381}{24}}+{b_{{2}}}^{3}+ \Bigl( {\frac {181}{18}}+b_{{3}} \Bigr) {b_{{2}}}^{2}
\nn\\
&\mbox{}+ \Bigl( {\frac {73}{18}}\,b_{{3}}+{\frac {4825}{126}}+{\frac {4}{21}}\,{b_{{3}}}^{2}
\mbox{} \Bigr) b_{{2}} \Bigr) {T}^{4}{\pi }^{4}{r_+}^{7}-157696\,K{T}^{3} \Bigl( {\frac {2131}{77}}+{b_{{2}}}^{3}+ \Bigl({\frac {5679}{616}}+{\frac {54}{77}}\,b_{{3}} \Bigr) {b_{{2}}}^{2}
\nn\\
&\mbox{}+ \Bigl({\frac {1299}{616}}\,b_{{3}}+{\frac {18087}{616}}+{\frac {3}{77}}\,{b_{{3}}}^{2}
\mbox{} \Bigr) b_{{2}} \Bigr) {\pi }^{3}{r_+}^{6}+258048\,K \Bigl(  \Bigl( MT\pi +{\frac {5}{32}} \Bigr) {b_{{2}}}^{3}+ \Bigl( {\frac {17}{21}}\,MT \Bigl( b_{{3}}+{\frac {562}{51}} \Bigr)\pi 
\nn\\
&+{\frac {251}{224}}-{\frac {5}{224}}\,b_{{3}} \Bigr) {b_{{2}}}^{2}
\mbox{}+ \Bigl( \frac{2}{21}\,M \Bigl( {b_{{3}}}^{2}+{\frac {157}{6}}\,b_{{3}}+{\frac {6937}{24}} \Bigr) T\pi -{\frac {67}{224}}\,b_{{3}}
\nn\\
&\mbox{}-\frac{1}{28}\,{b_{{3}}}^{2}+{\frac {601}{336}} \Bigr) b_{{2}}+\frac{3}{16}+{\frac {1793}{72}}\,MT\pi \Bigr){T}^{2}{\pi}^{2}{r_+}^{5}
\mbox{}-168960\,K T \Bigl(\Bigl( MT\pi -{\frac {27}{880}} \Bigr) {b_{{2}}}^{3}
\nn\\
&+\Bigl( {\frac {17}{55}}\, \Bigl( b_{{3}}+{\frac {1609}{68}} \Bigr) 
\mbox{}MT\pi -{\frac {63}{880}}\,b_{{3}}-{\frac {63}{176}} \Bigr) {b_{{2}}}^{2}
\mbox{}+ \Bigl( -{\frac {4}{55}}\,M \Bigl( {b_{{3}}}^{2}+\frac{3}{16}\,b_{{3}}-201 \Bigr)T\pi 
\nn\\
&-{\frac {63}{440}}\,b_{{3}}-{\frac {489}{440}} \Bigr) b_{{2}}-{\frac {57}{55}}+{\frac {961}{110}}\,MT\pi \mbox{} \Bigr) \pi \,{r_+}^{4}+153600\,K \Bigl(  \Bigl( {M}^{2}{T}^{2}{\pi }^{2}-{\frac {9}{1600}} \Bigr) {b_{{2}}}^{3}
\nn\\
&\mbox{}+ \Bigl( -{\frac {27}{800}}+{\frac {13}{25}}\,{M}^{2}{T}^{2} \Bigl( b_{{3}}+{\frac {363}{26}} \Bigr) 
\mbox{}{\pi }^{2}-{\frac {21}{100}}\,MT \Bigl( b_{{3}}+{\frac {75}{28}} \Bigr)\pi  \Bigr) {b_{{2}}}^2 
\nn\\
&+ \Bigl( -{\frac {27}{400}}+{\frac {39}{50}}\,{M}^{2}{T}^{2} \Bigl( b_{{3}}+{\frac {758}{39}} \Bigr) 
\mbox{}{\pi}^{2}-{\frac {147}{400}}\,MT \Bigl( b_{{3}}+{\frac {118}{21}} \Bigr) \pi  \Bigr) b_{{2}}
\nn\\
&-{\frac {9}{200}}-{\frac {369}{200}}\,MT\pi 
\mbox{}+{\frac {489}{50}}\,{M}^{2}{T}^{2}{\pi }^{2} \Bigr) {r_+}^{3}
\nn\\
&-27648\, \Bigl( b_{{2}}+\frac{3}{2} \Bigr) MK \Bigl(  \Bigl( MT\pi -\frac{1}{8} \Bigr) {b_{{2}}}^{2}+ \Bigl(-\frac{1}{2}-{\frac {7}{9}}\, \Bigl( b_{{3}}-{\frac {11}{7}} \Bigr)\mbox{}MT\pi  \Bigr) b_{{2}}
\nn\\
&-\frac{1}{2}-{\frac {11}{18}}\,MT\pi \mbox{} \Bigr) {r_+}^{2}+24576\, \Bigl( b_{{2}}+\frac{3}{2} \Bigr) ^{2}{M}^{2}K \Bigl(  \Bigl( MT\pi -\frac{3}{16}\Bigr) b_{{2}}+\frac{3}{2}\,MT\pi -\frac{3}{8} \Bigr)r_+
\nn\\ 
&+2048\, \Bigl( b_{{2}}+\frac{3}{2}\Bigr) ^{3}{M}^{3}K \Bigr]  \,.
\end{align}


\twocolumngrid

\bibliography{mybib}

\end{document}